\documentclass[11pt]{article}

\usepackage[margin=1in]{geometry}
\usepackage[T1]{fontenc}
\usepackage{lmodern}
\usepackage{amsmath}
\usepackage{amssymb}
\usepackage{mathtools}
\usepackage{microtype}
\usepackage{enumitem}
\usepackage[round,authoryear]{natbib}
\usepackage{hyperref}

\hypersetup{
  colorlinks=true,
  linkcolor=blue,
  citecolor=blue,
  urlcolor=blue
}

\newcommand{\doi}[1]{\href{https://doi.org/#1}{doi:#1}}
\newcommand{\PrE}{\Pr\nolimits_E}
\newcommand{\PrC}{\Pr\nolimits_C}
\newcommand{\Opre}{O_{\mathrm{pre}}}
\title{Does Probability Require a Single History?}
\author{Jonathan Baxter}
\date{October 2, 2026}

\begin{document}

\maketitle

\begin{abstract}
Everettian quantum mechanics is often said to lack coherent
probability because every possible measurement result occurs. If all
of them exist globally, what is probability supposed to be a
probability of? This paper defends an axiomatic answer: normalized
Born measure is postulated as objective probability over the records
borne by an observer's future continuations. Collapse interpretations
likewise postulate the Born probabilities, but instead assign them to
possible global outcomes. Neither theory derives the probabilities
from its ontology. The contribution is an explicit record-level probability
model that separates probabilities for future local records from certainty
about their global existence. I use this model to answer objections to the
axiomatic approach and argue that neither global selection nor a further
theory of personal persistence is required. At the level of recorded results,
the Everettian law supports
the same predictions and empirical inferences as the collapse law,
despite the theories' different ontologies.
\end{abstract}

\section{Introduction}
\label{sec:introduction}

Any interpretation of quantum mechanics must connect its formalism
with the Born probabilities. Everettian quantum mechanics is often
thought to face a special obstacle: while collapse interpretations
produce one measurement result, Everett contains a record of every
result represented with nonzero amplitude. If every result occurs, it
is natural to ask what a nontrivial probability could possibly
concern.

This paper takes the axiomatic route. It does not attempt to derive
probability from unitary dynamics or Born measure alone. Instead, it asks
what an explicit Everettian Born probability law would be a probability of.

Its central claim is simple. The probability need not concern whether a
particular record exists somewhere in the universal state: every record with
nonzero Born weight does. Instead, probabilities are assigned prospectively
to the records borne by an observer's future continuations. Each
continuation bears exactly one measurement record, although continuations
bearing every record with nonzero Born weight are actual. The relevant
uncertainty concerns the record value that will be local to a future observer,
not which records exist globally. The later observers, represented by their
physical observer-record states, thus form the domain of a future-record random
variable.\footnote{The construction uses the stable macroscopic records
identified by decoherence \citep{Zurek2003,Schlosshauer2005}.} The account then
postulates normalized Born measure as the objective probability distribution
of that variable. This postulate is not derived from unitary quantum
mechanics; it is added to it, just as a collapse theory adds a stochastic
Born law.

Here and throughout the paper, limited parity with collapse serves as a
diagnostic. The question in each case is whether the absence of collapse's
actualization of a single result creates an additional problem. If it does not, the issue is
common to both theories, and Everett therefore need not provide a deeper
explanation of what collapse itself postulates or assumes.

In ordinary measurement contexts that do not probe the coherence retained
between the resulting record states, the Everettian and collapse laws assign
the same Born probabilities to recorded results. They therefore make the
same predictions about what observers will record and support the same
evidential inferences from those records.\footnote{This record-level
correspondence does not make the theories physically or
ontologically equivalent. Collapse and Everettian quantum mechanics describe
different processes. A physical collapse suppresses or eliminates coherence
between outcome components that unitary evolution preserves, so sufficiently
sensitive interference or recoherence experiments can in principle distinguish
it from universal unitary evolution. Dynamical-collapse models such as GRW
make additional, model-specific
predictions by representing collapse as a precise stochastic modification of
the dynamics \citep{GRW1986}.} Collapse differs in making only
one measurement result globally actual. But such unique actualization
neither supplies the Born distribution nor is required for that distribution
to do its ordinary probabilistic work.

Existing Everettian accounts often combine a probability proposal with a
substantive theory of personal persistence, using overlapping persons,
temporal counterparts, worms, or threads to specify how subjects are related
across time. A recurring strategy recovers a one-to-one relation by positing
several physically indistinguishable premeasurement subjects, each with only
one future. This paper rejects the need for that reconstruction. The account
takes the one-to-many physical continuation relation at face value: each later
observer is a continuation of the one present observer. The
same local causal, physical, and psychological continuities that connect the
pre- and postmeasurement observer under collapse connect the present observer
to each later observer under Everett. Probability requires neither selecting
one of them as that observer's unique future nor positing a distinct present
antecedent for each future observer.

The contribution is to defend the sufficiency of the axiomatic approach:
an explicit Born law over future local records supports prediction and
empirical inference even when every record is globally actual. The formalism
keeps two questions separate: whether a record exists globally, and what
probability the law assigns to it as a future local record. A complete theory
of personal identity, a decision-theoretic derivation, or a reductive
metaphysics of chance may serve other aims, but none is required for the
probability law to perform these roles.

Section~\ref{sec:what-probability} develops the positive account. It
distinguishes global from local exclusivity; uses delayed revelation to
clarify local-record uncertainty; constructs a prospective sample space over
future continuations; states the Born law; and compares its predictions with
those of collapse. Section~\ref{sec:actualization} then considers objections
from selection, emergence, personal identity, and empirical confirmation,
including Adlam and Barandes's recent argument that the axiomatic route cannot
succeed \citep{AdlamBarandes2026}.
Section~\ref{sec:conclusion} concludes.

\subsection{The axiomatic route in the Everettian literature}

Everett used squared-amplitude measure to characterize the distribution of
records in repeated experiments. Although he described the resulting statistical assertions as
deducible from pure wave mechanics, the measure is naturally read as playing a
typicality role rather than as an explicit postulate of objective chance
\citep[p.~462]{Everett1957}.

Papineau defends treating squared amplitudes as objective probabilities even
when all outcomes occur, and argues that Everettian probability should not be
held to an explanatory standard that other accounts of physical chance do not
meet \citep{Papineau1996,Papineau2010}. Adlam explicitly rejects this comparison,
arguing that ordinary probabilities connect a theory to our particular
observations in a way that Everettian amplitudes do not
\citep[Section~IV.B]{Adlam2014}. The present paper addresses that objection by
specifying both what the Everettian probability law concerns and how its
probabilities enter empirical inference.

Ismael argues that Everettian outcomes can bear objective probabilities
despite deterministic unitary evolution
\citep[p.~437]{Ismael2011}. McQueen and Vaidman explicitly postulate
that squared amplitude gives the probability of self-location and compare
their postulate with the probability postulate required by collapse
\citep{McQueenVaidman2019}. They locate that uncertainty after measurement
and explain premeasurement behavior through concern for descendants rather
than premeasurement uncertainty.

Wilson develops objective Everettian probability through indexicalism, a
package combining divergent Everett worlds, individualism about possible
worlds, and indexical actuality \citep{Wilson2013}. The present account shares
his use of indexical structure to make sense of nontrivial objective
probability, but requires neither divergence nor the identification of Everett
worlds with metaphysically possible worlds.

Wilhelm's centered Everett interpretation likewise assigns objective Born
chances to propositions about an observer's local record, rather than whether
that record exists globally
\citep{Wilhelm2022,Wilhelm2023}. He develops this interpretation within a
worm ontology and a best-system account of chance. The present paper retains
his centered subject of probability without requiring that additional
structure.

Saunders and Wallace use overlapping persons to recover ordinary
premeasurement uncertainty \citep{SaundersWallace2008}. Barrett argues that
forward-looking probability requires suitable metaphysical scaffolding and
develops a many-threads account that supplies it \citep{Barrett2025}.

Tappenden proposes ``pure wave mechanics plus a no-collapse probability
postulate'' and applies it prospectively through a one-to-many account of an
observer's future \citep{Tappenden2021}. He embeds the proposal in a unitary
interpretation of mind, on which brains with identical cognitive contents
instantiate a single subject. The record-level model developed here does not
require those further metaphysical commitments.

\citet{Greaves2004} and \citet{Lewis2007} accept that a premeasurement observer
has several successors while denying that this supplies ordinary premeasurement
uncertainty. Greaves turns to a caring measure and decision theory, while Lewis
argues that an informed observer expects each successor and therefore lacks an
unknown result capable of grounding uncertainty. Greaves later argues that
empirical confirmation need not depend on resolving the dispute about
premeasurement uncertainty \citeyearpar{Greaves2007}. Her confirmation account
uses a decision-theoretic framework \citep[see also][]{GreavesMyrvold2010}.
The present account instead postulates an
objective-probability law for the mutually exclusive local-record values borne
by the later observers, without selecting a unique successor.\footnote{Even
if the decision-theoretic axioms are accepted
\citep{Deutsch1999,Wallace2012}, showing that rational agents
must use Born weights does not by itself establish that those weights are
the objective probabilities of their measurement records. Identifying the
rationally prescribed weights with objective chance must also account for
their role in measurement statistics
\citep[Section~3]{DawidThebault2014}. Here the objective-probability
interpretation is supplied explicitly by the Born postulate.}

\section{What Everettian probability is a probability of}
\label{sec:what-probability}

\subsection{Global and local exclusivity}

Consider an idealized two-outcome measurement. Before measurement, write the
state schematically as
\begin{equation}
  \lvert\Psi_{\mathrm{pre}}\rangle
  =
  \bigl(c_A\lvert A\rangle+c_B\lvert B\rangle\bigr)
  \lvert D_{\mathrm{pre}}\rangle
  \lvert \Opre\rangle
  \lvert E_{\mathrm{pre}}\rangle,
  \qquad |c_A|^2+|c_B|^2=1 .
  \label{eq:premeasurement}
\end{equation}
Here \(D\), \(O\), and \(E\) denote the apparatus, observer, and environment.
Under a collapse theory, measurement produces one of the possible record
states:
\begin{equation}
  \begin{aligned}
    \lvert\Psi_{\mathrm{pre}}\rangle
    &\xrightarrow{\text{collapse to }A}
    \lvert A,D_A,O_A,E_A\rangle, \\
    \lvert\Psi_{\mathrm{pre}}\rangle
    &\xrightarrow{\text{collapse to }B}
    \lvert B,D_B,O_B,E_B\rangle .
  \end{aligned}
  \label{eq:collapse}
\end{equation}
The outcome subscripts label the record encoded by the corresponding state.
A stochastic Born law says that the \(A\) transition occurs with probability
\(|c_A|^2\) and the \(B\) transition with probability \(|c_B|^2\).

By contrast, Everettian quantum mechanics retains both states:
\begin{equation}
  \lvert\Psi_{\mathrm{pre}}\rangle
  \xrightarrow{\text{unitary evolution}}
  c_A\lvert A,D_A,O_A,E_A\rangle
  +c_B\lvert B,D_B,O_B,E_B\rangle .
  \label{eq:everett-evolution}
\end{equation}
Environmental decoherence identifies those two states as stable macroscopic
record structures and makes interference between them negligible under
ordinary measurement conditions. It removes neither from the resulting state
and gives neither a probabilistic interpretation.

Collapse makes three claims coincide:
\begin{enumerate}[label=(\arabic*)]
  \item an \(A\)-record exists globally;
  \item the observer's local record is \(A\);
  \item the observer's future record will be \(A\).
\end{enumerate}
Within the \(A\)-collapse history shown in Equation~\eqref{eq:collapse}, all
three are true; within the \(B\)-collapse history shown there, all three are
false. The phrase
``the outcome is \(A\)'' can therefore stand for any of the three without
changing its truth value. Collapse thus supplies \emph{global} exclusivity:
within any resulting history, only one of the possible outcome records for
that measurement exists.

Everettian quantum mechanics does not supply global exclusivity: both \(O_A\)
and \(O_B\), bearing records \(A\) and \(B\), are present in the universal
state. It nevertheless supplies \emph{local} exclusivity at each later
observer. \(O_A\) bears \(A\) and not \(B\), while \(O_B\) bears \(B\) and not
\(A\). Here ``local'' means relative to one observer and their record, not
spatially local.

\subsection{Delayed revelation and local uncertainty}

A delayed-revelation experiment makes the distinction between global and local
exclusivity vivid. Suppose the observer is put to sleep before the measurement.
The result controls a transport mechanism, so the \(A\)-relative observer wakes in one room and the
\(B\)-relative observer in another. The rooms are internally identical and
each contains a sealed local record. The two observers occupy physically
distinct states correlated with different records, but have the same cognitively
available information. The record is revealed only later
\citep{Vaidman1998,McQueenVaidman2019}. The state during the delay has the form
\begin{equation}
  c_A\lvert A,D_A,O^A_{?},E'_A\rangle
  +c_B\lvert B,D_B,O^B_{?},E'_B\rangle .
  \label{eq:delayed-revelation}
\end{equation}
Writing \(i\) for either \(A\) or \(B\), the superscript in \(O^i_{?}\)
identifies the record physically local to the observer; the question
mark indicates that this value is not cognitively available. The prime marks
the environment state during the delay.

For each record value \(i\), compare the prerevelation observer in the possible
collapse history containing \(i\) with the actual Everettian observer
associated with \(i\). In both cases, record \(i\) is physically local to the
observer, but its value is not cognitively available. Knowing whether
the experiment involved collapse or unitary evolution changes what they know
about the global state, but does not reveal the local record. Delayed revelation
therefore establishes parity in the target of uncertainty: the hidden local
record is a coherent subject of uncertainty under either theory. Collapse
actualizes one such observer, whereas Everett instantiates one for every record
value. The existence of the other Everettian observers does not reveal the
local record to any one of them.

Once the result is revealed, each observer learns which record is in their
environment:
\begin{equation*}
  O^i_{?}\longrightarrow O_i .
\end{equation*}
The record value \(i\) is unchanged. The subscript in \(O_i\) indicates that
the record is now encoded in the observer's cognitive state as well as being
physically local.

Delayed revelation establishes the uncertainty parity just described, but is
not itself part of the probability model. The model starts from
\(\Opre\) and requires no actual prerevelation interval.

\subsection{The future-record variable}

I will make the distinction between a local record and its global existence
explicit by writing down a formal probability model for the records. We need
a sample space, a variable that returns the record at each sample point, and
a probability law.

Start with the premeasurement observer \(\Opre\). Let \(I\) be
the finite set of record values with nonzero Born weight. As before, \(O_i\)
denotes a later observer who knows that their local record is \(i\).
Decoherence identifies the stable local records that distinguish these
observers. Under unitary evolution, each \(O_i\) continues from
\(\Opre\):
\begin{equation}
  \Opre
  \longrightarrow
  \{O_i:i\in I\}.
  \label{eq:continuations}
\end{equation}

The arrow in Equation~\eqref{eq:continuations} marks the direction of stable
record formation, not a fundamental irreversibility in the unitary
dynamics.\footnote{Explaining this macroscopic arrow therefore requires more
than unitary dynamics alone. Low initial entanglement has been proposed as a
special cosmological boundary condition that could supply the asymmetry,
perhaps in connection with the low gravitational entropy of the early
universe and the corresponding thermodynamic arrow
\citep{AlKhaliliChen2024}.}

Compare \(O_i\) with the observer in the collapse history containing record
\(i\). At the coarse-grained record level, they have the same physical
observer-record state and the same relevant causal, physical, and
psychological continuity with \(\Opre\). Collapse makes only one
of these later observers actual; Everett makes them all actual. I take each
\(O_i\) to be a future continuation of the present observer in either case.
Section~\ref{sec:personal-continuation} defends this claim.

When I say that an observer knows their record, I mean that the record is
encoded in their physical state and is cognitively available. Like
\citet[pp.~234--236]{Papineau1996}, I assume that such a state supports
conscious experience on the same basis under Everett as under collapse.
Whatever makes the postcollapse observer aware of record \(i\) also makes
the corresponding Everettian observer aware of it. This is an interpretive
assumption, not something established by the probability law.

The model uses only these coarse-grained observer-record distinctions.\footnote{
In particular, the construction uses the later observers \(O_i\), distinguished
only by their coarse-grained physical record states, not an inventory of
branches. A more detailed microscopic description may resolve the same
observer-record state into several orthogonal components. These introduce no
additional record value and do not change the total Born weight assigned to
the record.
Branch counting instead treats the number of branches bearing each record as
relevant to rational credence, but then faces the difficulty that the
branches to be counted are not canonically determined by the underlying
quantum dynamics. For a recent treatment of this problem, see
\citeauthor{Khawaja2026}'s (\citeyear{Khawaja2026}) observation-indexed
proposal. Assessing such proposals lies beyond the present argument because
nothing in the probability model developed here depends on a branch count.}
Its sample space is the set of later observers:
\begin{equation}
  \Omega_E=\{O_i:i\in I\}.
  \label{eq:everett-space}
\end{equation}
Calling this a sample space does not mean that only one of its elements
exists. It is the mathematical domain of the probability law. Its elements
are the later continuations of this particular \(\Opre\), so the
probabilities concern that observer's future records.

Define the Everettian future-record random variable by
\begin{equation}
  X_E:\Omega_E\longrightarrow I,
  \qquad X_E(O_i)=i .
  \label{eq:record-variable}
\end{equation}
The sample points are observers; the values of \(X_E\) are their records.
In the two-outcome example, \(X_E\) returns \(A\) at \(O_A\) and \(B\) at
\(O_B\). Both observers exist, but each bears only one record. The variable
therefore has one value at each observer, not both values at either.

From \(\Opre\)'s point of view, each of these later observers is
a future continuation. But they do not share a single experience of all the
records. Each knows one record as their own. The present observer knows that
every record will occur, but does not yet bear any of those future local
records. It is these differing future local-record values that the
prospective uncertainty concerns, not which observer will exist or which
one will uniquely be \(\Opre\). In this sense the uncertainty is
future-centered and non-singular: it concerns the local records at several
future continuations of one present observer.

The continuation relation identifies the future local-record values at issue
but does not determine their numerical probabilities. The normalized Born
measure \(\mu_{\Psi}\), introduced next, supplies the measure on \(\Omega_E\)
needed to complete the mathematical definition of \(X_E\). The Born postulate
then interprets that measure as objective probability.

\subsection{The Born probability law}

The quantum state already gives a squared-amplitude measure to the record
alternatives. In the two-outcome example, their normalized weights are
\(|c_A|^2\) and \(|c_B|^2\). More generally, if \(c_i\) is the amplitude of
the component bearing record \(i\), write
\begin{equation}
  \mu_{\Psi}(X_E=i)=|c_i|^2,
  \qquad \sum_{i\in I}|c_i|^2=1 .
  \label{eq:born-measure}
\end{equation}
Here \(\mu_{\Psi}\) is the Born measure fixed by the quantum state.
For a macroscopic record, \(|c_i|^2\) is the total squared
amplitude of the components bearing that record. We can also represent the
record subspace by a projector \(\Pi_i\), in which case its weight is
\(\langle\Psi\rvert\Pi_i\lvert\Psi\rangle\). The simpler amplitude notation
will suffice here.

Equation~\eqref{eq:born-measure} fixes the weights, but does not yet say
that they are objective probabilities. That is the additional postulate
defended here:
\begin{equation}
  \boxed{
  \PrE(X_E=i)
  =\mu_{\Psi}(X_E=i)
  =|c_i|^2 .}
  \label{eq:everett-postulate}
\end{equation}
This is the Born probability postulate, not a theorem of unitary quantum
mechanics. The probabilities concern the local-record values returned by
\(X_E\), not the existence of the observers who bear them.

Does the global coexistence of all the later observers rule out this law? Not
by itself. Consider a playing-card analogy. The four aces from a deck are
shuffled fairly and dealt one to each of four participants. Every ace is
certain to be dealt. But ``the ace of spades will be dealt'' and ``I will
receive the ace of spades'' are different events: the first has probability
one, the second probability \(1/4\). All four aces are dealt, but each
participant receives only one.

The analogy shows how every result can occur while each player's own
result remains a target of objective probability. It still involves one actual
deal, however, and the players are already distinct before it. Under Everett,
the later observers all continue the same present observer. But the players' prior separateness is
not an ingredient of the card probability assignment. The assignment concerns
the card each receives under the fair-dealing law. The Everettian construction
likewise concerns the record each later observer bears, with its probability
supplied by the Born postulate. A critic who rejects this law because all the
observers are actual must explain why objective probability requires global
exclusivity in addition to the local exclusivity supplied here.

Reconstructing several distinct premeasurement subjects would nevertheless
make the situation more familiar. Each subject would resemble a player
awaiting their card: there is one future result that is theirs, although they
do not know which. Such a reconstruction restores the ordinary singular form
of uncertainty without introducing collapse, but the probability law does not
require it.

Without that reconstruction, however, ordinary first-person grammar makes the
simple probability law in Equation~\eqref{eq:everett-postulate} difficult to
express. We can say, precisely but awkwardly,
``The future-local-record variable defined over the continuations of the
present speaker has value \(A\) with probability \(p\).'' The natural
alternative---``I will record \(A\) with probability
\(p\)''---presupposes that the present speaker has one future bearer and
invites the question, ``Which future continuation will be me?''%
\footnote{Ismael likewise argues that a present first-person thought in an
Everettian universe does not pick out a unique future trajectory
\citep[pp.~786--789]{Ismael2003}.}

Expressing this one-to-many relation while keeping the present speaker as the
grammatical subject would require a future-plural first-person pronoun. For
expository purposes, I will use \emph{we-I}. ``We-I will record \(A\) with
probability \(p\)'' is shorthand for
\begin{equation*}
  \PrE(X_E=A)=p.
\end{equation*}
We can describe the same continuation relation without the new pronoun:
``My future continuations will each record one outcome.'' But that wording
puts the continuations at a distance. It can make them sound like later
observers related to me rather than future versions of me. We-I keeps the
claim in the first person: ``We-I will each record one outcome.''

The pronoun refers separately to each future observer, not to a single
experience shared by them all. It posits neither a group mind nor
several hidden observers already present before measurement.

We-I will therefore record \(i\) with probability \(|c_i|^2\).

\subsection{The correspondence with collapse}

A collapse theory assigns the same Born probabilities to the possible
recorded results, but over a different sample space. In the two-outcome model
of Equation~\eqref{eq:collapse}, write
\(G_i:=\lvert i,D_i,O_i,E_i\rangle\), \(i\in\{A,B\}\), for the possible global
states after collapse. For a general record set \(I\), define
\begin{equation}
  \Omega_C=\{G_i:i\in I\},
  \qquad
  X_C:\Omega_C\longrightarrow I,
  \qquad X_C(G_i)=i .
  \label{eq:collapse-space}
\end{equation}
Thus \(X_C\) is the random variable that returns the unique record in each
possible postcollapse global state. Its stochastic law is
\begin{equation}
  \PrC(G_i)
  =\PrC(X_C=i)
  =|c_i|^2 .
  \label{eq:collapse-postulate}
\end{equation}
Both laws give record value \(i\) probability \(\lvert c_i\rvert^2\). However,
their sample spaces differ. Under collapse, a sample point \(G_i\) is a
possible postcollapse global state, and \(X_C\) returns the record it contains.
Under Everett, a sample point \(O_i\) is one of the actual later observers
continuing the present observer, and \(X_E\) returns the record local to that
later observer.
The variables \(X_E\) and \(X_C\) therefore have different domains but the same
possible values and probability distribution. Thus
\begin{equation}
  X_E\overset{d}{=}X_C .
  \label{eq:equal-distribution}
\end{equation}
This is a record-level probabilistic correspondence, not an isomorphism between
the theories as a whole.

The single-trial law extends by independence. For \(n\) repetitions, write
\(\mathbf i=(i_1,\ldots,i_n)\) for a record sequence, and let \(X_{E,m}\)
return the record of trial \(m\). The joint distribution is
\begin{equation}
  \PrE(X_{E,1}=i_1,\ldots,X_{E,n}=i_n)
  =\prod_{m=1}^{n}\PrE(X_{E,m}=i_m)
  =\prod_{m=1}^{n}|c_{i_m}|^2.
  \label{eq:sequence-probability}
\end{equation}
The collapse law assigns the same product distribution.

Returning to the two-outcome case \(I=\{A,B\}\), let \(K\) be the number of
\(A\)-records in the sequence. Either law gives
\begin{equation}
  \Pr(K=k)=\binom{n}{k}|c_A|^{2k}|c_B|^{2(n-k)} .
  \label{eq:binomial}
\end{equation}
Under collapse, this distribution ranges over \(2^n\) possible global
histories, only one of which becomes actual. Under Everett, the same \(2^n\)
record sequences distinguish \(2^n\) actual later observers at the
coarse-grained level of the model. In this context, when
\(\Opre\) uses we-I, the pronoun refers to these future
continuations, each bearing its own record sequence.
Equation~\eqref{eq:sequence-probability} gives the prospective probability
distribution over those sequences. By the law of large numbers, as \(n\)
increases, we-I will record an \(n\)-trial sequence whose relative frequency of \(A\) is close to
\(|c_A|^2\), with probability approaching one. Here it is the total Born
probability of sequences with such frequencies that approaches one, not
necessarily the fraction of all sequences they constitute. Every sequence
remains actual, but coexistence does not imply equal probability.

The same distribution supplies the numerical likelihoods used in the
next section. The Born law assigns an observed record sequence
\(\mathbf i\) its usual likelihood, not likelihood one merely because
that sequence is instantiated somewhere globally.
Section~\ref{sec:evidence} compares Born measure with uniform counting
and explains how these record likelihoods support empirical
confirmation.

\section{Does probability require unique actualization?}
\label{sec:actualization}

The central probability construction has now been stated: decoherence identifies
stable physical observer-record states; the one-to-many continuation relation
identifies the observers they support as later continuations of
\(\Opre\); \(X_E\) returns their mutually exclusive local records;
and the explicit Born law assigns probabilities to those records. It remains to
ask whether the construction tacitly depends on additional structure and
whether its probabilities can support empirical confirmation.

I organize those questions around Adlam and Barandes's
recent paper, which brings together several of the principal objections and
makes an unusually strong claim: axiomatic, deductive, and inductive approaches
all face fundamental obstructions \citep{AdlamBarandes2026}. If their argument
succeeds, it rules out the present account. I argue that it does not.

Because Adlam and Barandes frame these objections in terms of branches, I
retain that language below. Nothing turns on this choice, since the objections
apply equally to the record-level construction developed above. I address them
under four headings: global selection, emergent records, one-to-one personal
continuation, and confirmation.

\subsection{Selection and exclusivity}
\label{sec:selection}

Adlam and Barandes compare branch weights with the proportions of differently
colored marbles in a jar. If 43 of 100 marbles are red, the fraction of red
marbles is \(0.43\), but it is not yet the probability of anything. A random
drawing or some other selection process must be specified before \(0.43\)
can be interpreted as the probability of drawing red. Likewise, they argue,
the mere coexistence of branches with different weights supplies no
probability unless something selects one branch.

They are right that a measure over coexisting objects is not by
itself a probability, so Equation~\eqref{eq:born-measure} does not entail
Equation~\eqref{eq:everett-postulate}.
Equation~\eqref{eq:everett-postulate} supplies the missing step axiomatically:
it interprets Born measure as probability. Their marble analogy does not,
however, show that such a law must describe a global selection. Here the law
applies to \(X_E\), which returns the record borne by each later observer
continuing \(\Opre\), not to the global existence of those records.

Collapse and Everett provide exclusivity differently. Collapse leaves a
single record in the global state. In the Everettian sample space, every
later observer is actual, but \(X_E\) has exactly one value at each.
Adding a process that selected one record as the unique global result would
not show that \(X_E\) is ill-defined. It would replace \(X_E\) with the
collapse variable \(X_C\).

A collapse advocate might reply that this is precisely the problem: collapse
supplies genuine premeasurement possibilities, whereas every Everettian future
observer is actual. But treating this difference as disqualifying assumes
the point at issue: that probability must concern global possibilities of which
only one becomes actual. On the account developed above, the probability is
prospective relative to \(\Opre\). Each of that observer's later
continuations experiences one local record, not all the records together, and
the law assigns probabilities to those locally exclusive values.

Part of the intuitive difficulty is that equally definite, equally actual
experiences coexist, yet their records receive unequal probabilities. One
later observer experiences record \(A\); the other experiences record \(B\).
Neither experience is less definite or less real because its record has a
lower Born weight. This apparent subjective symmetry can make unequal
probabilities seem puzzling.

Collapse may be appealing in part because it removes the coexistence that
makes this contrast troubling. Only one result becomes actual, so there are
no coexisting observers whose equally definite experiences invite the
comparison.

But selection does not determine the Born distribution. A process that
produces one global result could follow any probability law. Collapse assigns
that result probability \(\lvert c_i\rvert^2\) only through its stochastic Born
postulate. The account developed here makes the parallel move for \(X_E\): its
law has a different domain but postulates the same distribution. The two
questions remain distinct: what probability concerns and what probabilities
the law assigns.

\subsection{Emergent records and Born measure}

Adlam and Barandes also argue that a fundamental axiom cannot assign a new
primitive property to an emergent object. Their analogy is an axiom declaring
that tables and chairs derived from an underlying physical theory must be
yellow. Because branches and observers emerge through decoherence, they
conclude that probabilities attaching to them must be derived from the
fundamental physics rather than postulated.

A chair's color, however, is a physical feature constrained by the
underlying state and dynamics. The Born postulate does not change the
physical records obtained from unitary dynamics. It supplies a
probability interpretation for their weights that those dynamics do
not themselves provide.

Although the records are emergent, their Born weights are fixed by the
underlying quantum state. Unitary measurement dynamics correlates each
microscopic alternative with a macroscopically distinct apparatus and
observer record while preserving its amplitude. Environmental
decoherence makes the resulting record structures effectively
orthogonal. The Born weight of a macroscopic record is the squared
norm of the corresponding part of the quantum state, not an additional
quantity supplied by the probability postulate.

Adlam and Barandes also argue that adding fundamental branches
undermines Everett’s claim to require no additional mathematical
machinery beyond unitary quantum mechanics. But a probability law can
refer to derived record structures without making them fundamental
objects. Here the record projectors \(\Pi_i\) represent stable
subspaces identified by the measurement interaction and environmental
decoherence \citep{Zurek2003}. The Born postulate is an additional
commitment: it interprets the weights of those subspaces as objective
probabilities. It introduces neither fundamental branches nor new dynamics.

A chair does provide a useful analogy for the issue of
precision. There is no perfectly sharp fact about which particles
throughout its surface region belong to it, so its exact mass is
correspondingly vague. But that does not make mass inapplicable at the
macroscopic scale. Likewise, uncertainty at the boundaries of a record
structure may make its precise Born weight approximate without making
Born weight inapplicable.

There are well-known structural reasons to privilege squared amplitude
over other ways of assigning weights to the record alternatives. Born
weight is additive across orthogonal record components and invariant
under changes of basis within a fixed record subspace. Gleason-style
results make the uniqueness claim precise under suitable assumptions
\citep{Gleason1957}.  Such results explain why Born measure is the
natural measure supplied by the quantum formalism, but they do not
turn that measure into objective probability.  That final
interpretation is exactly what Equation~\eqref{eq:everett-postulate}
states.

Collapse also requires a probability postulate. A collapse theory may
build the Born values into its stochastic dynamics, but their
probabilistic status is still part of the theory's law. Unique
actualization supplies neither those values nor a reductive analysis
of objective chance. Everettian quantum mechanics need not derive what
collapse also postulates.

\subsection{One observer, several future continuations}
\label{sec:personal-continuation}

A further worry is that \(X_E\) cannot represent what the present
observer will record unless each future observer has a distinct
premeasurement antecedent with only that future. Some Everettian
responses accept this demand, using overlapping or divergent persons,
spacetime worms, or nonbranching threads to recover a one-to-one
structure
\citep{SaundersWallace2008,Wilhelm2022,Barrett2025}.\footnote{Lewis
considers this one-to-one strategy as one of several accounts of
branching selves, but argues that it cannot provide uncertainty
capable of grounding probability \citep{Lewis2007}.}  Each future
observer then has its own physically indistinguishable premeasurement
antecedent, which may be uncertain which future is theirs. This
restores a familiar model of uncertainty and probability, but accounts
constructed this way must defend both the probability proposal and the
accompanying theory of personal persistence.
  
Barrett makes the underlying concern explicit. If the present observer
regards both later observers as future selves, he argues, each result
should receive probability one. If neither later observer is identical
to the present observer, each result should receive probability zero
\citep[pp.~66--68]{Barrett2025}.

The first conclusion does not follow for the future-record
variable. The present observer is certain that each result will be
recorded by a future continuation, but that is not the claim that each
value of \(X_E\) has probability one. In the two-outcome model, the
event \(X_E=A\) contains \(O_A\), not \(O_B\), even though both are
actual future continuations. The Born law assigns probabilities to
these distinct local-record events, not to the existence of the
observers bearing them.

Nor does rejecting numerical identity require assigning probability
zero.  Not every theory of personal persistence imposes a one-to-one
structure.  \citet[p.~237]{Papineau1996} appeals to Parfit's notion of
personal survival to accommodate one earlier self and several later
selves.  Tappenden's account retains a one-to-many relation within a
theory of temporal counterparts \citep{Tappenden2021}. The account
developed here likewise takes personal continuation to be
one-to-many. Each \(O_i\) is genuinely a future continuation of
\(\Opre\), without being numerically identical to it:
otherwise, transitivity would make the later observers identical to
one another.

The positive case for this continuation relation rests on parity with collapse.
For each record \(i\), the corresponding postmeasurement observer has
the same relevant causal, physical, and psychological continuity with
\(\Opre\) under collapse and unitary evolution. The same grounds for
regarding the collapse observer as a future version of \(\Opre\)
therefore apply to the corresponding \(O_i\) under Everett.
  
A collapse theorist might insist that a later observer continues the
present one only if no alternative outcome is actual. That would make
continuation depend not just on the relation between the earlier and
later observer, but also on whether other outcomes occur. This additional
requirement needs independent justification. It cannot simply be assumed
in an argument intended to show that probability requires unique
actualization.

In the Everettian model, the continuation relation identifies the later
observers that form the sample space \(\Omega_E\), and their records
determine the values of the future-record variable \(X_E\). The Born
postulate supplies the probability distribution. A further theory of
personal persistence that preserves these continuations and their records
leaves all three unchanged. It may answer other questions about personal
identity, but it is not needed to complete the probability model.

\subsection{Evidence when every result occurs}
\label{sec:evidence}

Adlam and Barandes argue that Everettian quantum mechanics cannot
support inductive expectations of Born-rule regularities because every
possible future record sequence occurs. The universal state follows a
consistent unitary law, but, they argue, this supplies no basis for
expecting regularity in an observer's future records
\citep[Section~7]{AdlamBarandes2026}.

If, as they assume, the axiomatic and deductive approaches have been ruled
out, then they are right: there is no probability law to support inductive
expectations. Much of their subsequent argument depends on that exclusion.
But the preceding sections have defended an explicit probability law and
explained why their objections to the axiomatic approach fail.
The relevant question here is therefore whether the probability law already
defended supports inductive expectations and empirical inference. It does.

For independent repetitions of the two-outcome measurement, the Born law
gives the binomial distribution in Equation~\eqref{eq:binomial}.
For any fixed positive tolerance,
the total probability of record sequences whose relative frequency of \(A\) differs
from \(|c_A|^2\) by more than that tolerance tends to zero as the number of
repetitions \(n\) increases.
Evidence supporting the Born law therefore also supports expecting Born-rule
regularities in future records.

Yet Adlam and Barandes claim that ``only very rare branches have the right
kind of internal uniformity that conforms to the Born rule'', contrasting these with the
``vast majority'' of branches that lack such regularity
\citep[p.~15]{AdlamBarandes2026}. They do not specify a measure, but the
comparison is between how many histories exhibit Born-rule regularities
and how many do not. For record sequences, this corresponds to counting
each sequence equally, regardless of its Born weight, which is equivalent
to using the uniform (normalized counting) measure. For binary trials,
the uniform measure is the product measure with equal single-trial weights
of 0.5 for \(A\) and \(B\), which coincides with Born measure for independent
equal-amplitude preparations.

Now consider \(n\) independent repetitions of a binary measurement with unequal
Born weights, say 0.9 for \(A\) and 0.1 for \(B\). Under the uniform measure,
sequences with frequencies close to the Born values---for example, an
\(A\)-frequency between 0.85 and 0.95---constitute a vanishing fraction of the
\(2^n\) sequences as \(n\) increases, consistent with their claim of rarity.
The total Born weight of those sequences, on the other hand, approaches one
by the same law of large numbers argument given above.
Under the Born postulate, these sequences are therefore overwhelmingly
probable. Nothing about the coexistence of all the sequences requires a uniform
measure over them, despite the intuitive appeal of equating equal actuality with
equal probability (Section~\ref{sec:selection}). Rarity under that measure does
not undermine inductive expectations based on the Born law.

Adlam and Barandes further argue that rescuing induction would require an axiom
stipulating that our past and future records conform to the Born
rule. Once again, this assumes that the explicit Born postulate has
been ruled out. With that postulate, Born-rule regularities are highly
probable, not inevitable, and the same law applies to every observer.

The coexistence of every sequence also does not prevent empirical
testing. Two candidate laws can agree that every sequence occurs while
assigning different probabilities to those sequences as future local
records. An observer's evidence is the sequence they actually record,
not the fact that it exists somewhere. The probabilities each law
assigns to that sequence supply the likelihoods in Bayes' rule,
allowing the observer to update their probabilities for the laws and,
consequently, their predictions for future records.

Finally, they argue that adding the Born rule as a brute postulate does
not explain the observed statistics. But explaining those statistics is different
from explaining the probability law itself. A collapse theory also postulates
a stochastic law. Saying that collapse selects an outcome with its Born
probability states that law rather than explains it. Suppressing or eliminating
the other outcome components adds a physical change, not an explanation of
why the probabilities have those values.

In both accounts, the probability law provides a statistical explanation by
assigning high probability to Born-rule regularities. As shown here, the
Everettian law can do that without a physical process selecting one globally
actual result. Its being postulated rather than derived is therefore no special
explanatory deficit for Everett. In either case, the theory together with its
probability law is tested against the records, not assumed correct in advance.

\section{Conclusion}
\label{sec:conclusion}

The familiar probability objection begins with a true global claim: in
Everettian quantum mechanics, every measurement record with nonzero amplitude
exists. It then asks what a nontrivial probability could concern if every
record occurs. Collapse makes it natural to frame probability as the question
``Which record will occur?'' because global occurrence, the observer's local
record, and the observer's future record coincide. Under Everett
they do not.

On the account defended here, an Everettian probability law does not assign
probabilities to whether records exist globally. It assigns them prospectively
to the records borne by the observer's future continuations. Delayed revelation
establishes that the hidden local record is the same target of uncertainty
under collapse and Everett. The one-to-many continuation relation
supplies the later observers, the future-record variable returns their records,
and normalized Born measure supplies the corresponding weights. The
Born probability postulate interprets those weights as objective
probabilities:
\[
  \PrE(X_E=i)=|c_i|^2.
\]

The corresponding collapse variable has the same distribution. Consequently,
the two laws make the same record-level predictions, generate the same
frequency expectations, and assign the same likelihoods to observed evidence.
Their agreement at that level does not erase their different dynamics or
ontologies.

The construction assumes that decoherent observer-record states support the
corresponding conscious perspectives under Everett just as the analogous
postcollapse states do. Given that premise, a complete theory of personal
identity, a decision-theoretic account of rational action, and a reductive
theory of objective chance may serve broader aims, but none is required for
this record-level probability law to support prediction and empirical inference.

The question in the title can be asked at two levels. Must one history be
selected as uniquely
actual? Or, without such a selection, must the present observer be redescribed
as several distinct subjects, each confined to one history, so that the
familiar one-to-one picture of personal persistence can be preserved? The
answer to both is no.
Probability can concern the local record borne by a later observer even
though, for every allowed record, a later observer bearing that record is
actual. And on the account developed here, every later record-bearing observer
is a continuation of the one present observer.
Probability therefore requires neither a uniquely actual global history nor a
distinct single-history antecedent for each future observer.

\end{document}